\documentclass[letter,twocolumn]{jpsj3} 
\usepackage{bm,amssymb,amsmath}
\usepackage{graphicx}   
\usepackage{txfonts}
\usepackage{braket}
\usepackage[table]{xcolor}
\usepackage{arydshln}
\usepackage{mathrsfs}
\makeatletter
\newcommand{\figcaption}[1]{\def\@captype{figure}\caption{#1}}
\newcommand{\tblcaption}[1]{\def\@captype{table}\caption{#1}}
\makeatother

\title{
RKKY Interactions and Multipole Order in Ab initio Wannier Model of CeCoSi
}

\author{Takemi Yamada\thanks{E-mail address: t-yamada@pu-toyama.ac.jp}, Yuki Yanagi and Keisuke Mitsumoto}
\inst{
Liberal Arts and Sciences, Toyama Prefectural University, Imizu, Toyama 939-0398, Japan}
\abst{
We calculate the RKKY interactions derived from ab initio calculations for the intermetallic compound CeCoSi exhibiting the hidden nonmagnetic order at $T_0$ and examine the instability towards possible multipole orders within the random phase approximation. All 36 multipole interactions up to rank 5 are investigated, and the maximum susceptibility exhibits an antiferro order with $\boldsymbol{q}=\boldsymbol{0}$ for nonmagnetic multipoles of monopole $I$ and hexadecapole $H_{0}$, yielding a charge imbalance of $f$ electrons at two Ce atoms in the unit cell. The obtained order can explain some experiments. 
}

\begin{document}
\maketitle
The tetragonal intermetallic compound CeCoSi exhibits the so-called hidden order~(HO) at slightly higher temperature $T_0\simeq 12$~K\cite{Tanida2018,Tanida2019} than the antiferromagnetic~(AFM) order $T_{\rm N}=9.4$~K. The absence of magnetic order with temperatures $T_{\rm N}<T<T_0$\cite{Manago2021} has attracted much attention in terms of non-magnetic higher-order multipole orderings, including cluster multipole orderings\cite{YH-JPSJ2020,YH-PRB2020}. In the normal phase $T>T_0$, 4$f$ electrons show the localized behavior in bulk experiments\cite{Tanida2019,Hidaka2022}, and the crystalline electric field (CEF) level splits into three Kramers doublets~(KDs) with two CEF excitation energies around 100-150~K determined by the inelastic neutron scattering\cite{Nikitin2020}. 
The fact that the nonmagnetic HO occurs first is interesting, since at very low temperatures only the lowest KD is active, and the onset of magnetic order is generally expected. It is also noteworthy that the pressure $P$-dependence of $T_0$ shows a dome-shaped pressure phase diagram similar to the so-called Doniach phase diagram, increasing to about 40~K at $P\simeq 1.5$~GPa and disappearing at $P\simeq 2.2$~GPa\cite{Tanida2019}, reminding us of the quantum critical point of the nonmagnetic multipole. 

Recent X-ray diffraction experiment has revealed that the electronic state at $T<T_0$ is accompanied by a structural transition to a triclinic structure\cite{Matsumura2022}, and the interpretation of $T_0$ as a ferro-quadrupole order of $O_{zx}+O_{yz}$ is also supported by NMR measurement\cite{Manago2023}. The magnetic phase diagram\cite{Hidaka2022,YH2022} at low temperatures including structural transitions is very complicated depending on the direction of the magnetic field, and is still being studied intensively toward a unified understanding. However, the proposed quadrupole mechanisms\cite{YH-JPSJ2020,YH-PRB2020,Manago2021,YH2022} require the quadrupole interactions to be large enough to overcome the lowest KD fluctuations that increases at low temperatures, but the magnitude and sign of the interactions and their type dependence are only assumed phenomenologically. Therefore, the microscopic analysis of multipole interactions reflecting the density functional theory~(DFT)-based bandstructure, fully including the conduction ($c$) bands and the mixing between localized $f$ and itinerant $c$ electrons~($c$-$f$ mixing), is crucial for the theoretical clarification of HO. 

In this letter, we report a study of the Ruderman-Kittel-Kasuya-Yosida~(RKKY) interactions based on the multi-orbital Wannier model derived from the ab initio calculation on CeCoSi, and the possible multipole order within the random phase approximation~(RPA). This approach is an extension of our work on the typical localized multipole system CeB$_6$\cite{YH2019,YH2020} and should be useful for understanding CeCoSi.

\begin{figure}[t]
\centering
\includegraphics[width=8.0cm]{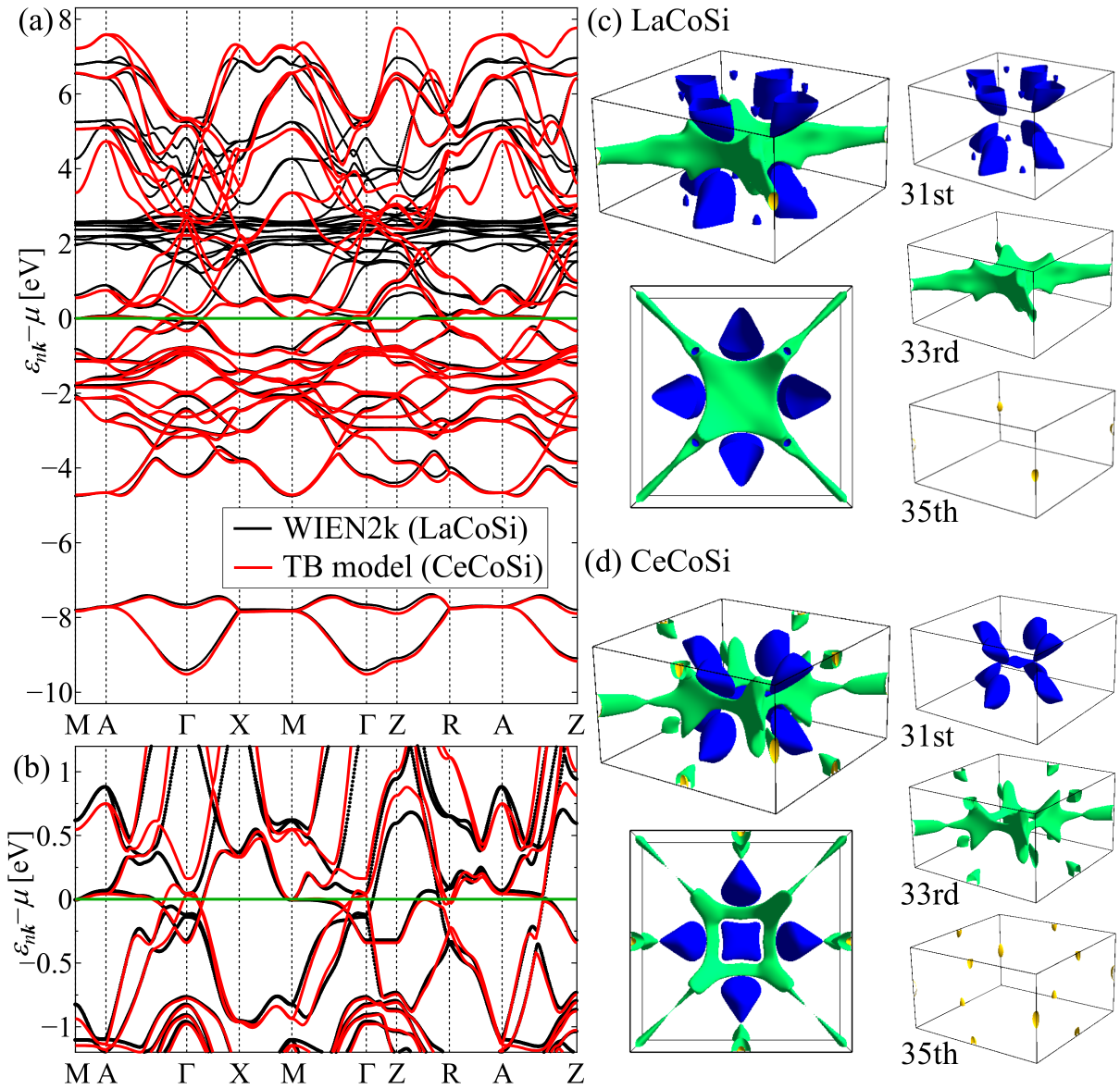}
\vspace{-2mm}
\caption{(Color online) 
(a),(b) The $c$ bands of TB model~(red) and the DFT bands of LaCoSi~(black) in (a) wide and (b) narrow energy ranges, where 
horizontal green line is Fermi-level and the high-symmetry points are $\Gamma~(0,0,0)$, X~$(\pi,0,0)$, M~$(\pi,\pi,0)$, Z~$(0,0,\pi)$, R~$(\pi,0,\pi)$ and A~$(\pi,\pi,\pi)$. (c),(d) The corresponding FSs for (c) LaCoSi and (d) CeCoSi depicted by Fermisurfer\cite{Fermisurfer2019}. 
}
\label{Fig01}
\end{figure}

First we perform band calculation for CeCoSi with the experimental lattice parameters by using the DFT-based first-principles package WIEN2k\cite{Blaha2020}, and then construct a 84-orbital tight-binding~(TB) model from the Wannier~90\cite{w90-Pizzi2020} consisting of Ce-$4f,5d$, Co-$3d$ and Si-$3p,3s$ orbitals together with the sub-lattice due to 2CeCoSi/unit-cell~(uc) and spin degrees of freedom. The computational details are presented in \S1 and \S2 of Supplemental Material~(SM)\cite{SM}. 

The obtained TB Hamiltonian is given by the following, 
\begin{align}
\mathscr{H}_{\rm TB}=\mathscr{H}_{c\textrm{-}c}+\mathscr{H}_{f\textrm{-}f}+\mathscr{H}_{c\textrm{-}f}
\end{align} 
where $\mathscr{H}_{c\textrm{-}c}~(\mathscr{H}_{f\textrm{-}f})$ is the $c$-$c$~($f$-$f$) term for $c~(f)$ electrons with 56~(28) orbitals and $\mathscr{H}_{c\textrm{-}f}$ is the $c\textrm{-}f$ mixing term. By diagonalizing $\mathscr{H}_{c\textrm{-}c}$ in the wavevector $\bm{k}$ space, we can obtain the $c$ band dispersion $\varepsilon_{n\bm{k}}$ with band-index $n$. 

Figures~\ref{Fig01}(a) and (b) show $\varepsilon_{n\bm{k}}-\mu$ of the TB model for CeCoSi together with the DFT bandstructure for LaCoSi in wide [narrow] energy range as shown in Fig.~\ref{Fig01}(a) [Fig.~\ref{Fig01}(b)], where the chemical potential $\mu$ is determined to satisfy the $c$ electron number $n_{c}=32$. The two bands are similar especially for the occupied states below Fermi energy~($\varepsilon_{n\bm{k}}=\mu$), which indicates that $\mathscr{H}_{c\textrm{-}c}$ can well describe the $c$ electron state of LaCoSi almost without $f$ electrons and also is consistent with the angle-resolved photoemission spectroscopy~(ARPES) experiment on CeCoSi, where the DFT bands of LaCoSi suggests relatively good correspondence with the ARPES\cite{Kimura2021}. 

\begin{table}[t]
\caption{
Classification of 36 multipole operators by IRR in the point group $C_{4v}$ and TR symmetry~\cite{Kusunose2008,Hayami2018,Ikeda2012}, where even~(odd) rank multipoles of monopole (dipole), quadrupole (octupole) and hexadecapole (triakontadipole) are denoted by $I~(J)$, $O~(T)$, and $H~(D)$, respectively\cite{Ikeda2012}~(see \S3 of SM\cite{SM}).
}\label{table01}
\vspace{1mm}
\scalebox{0.85}{
\begin{tabular}{ccc}
\hline \rule{0pt}{4mm}
IRR & TR-even & TR-odd \\[2pt] \hline\hline \rule{0pt}{4mm}
$\Gamma_{1}~(A_{1})$ & $I,~O_{u},~H_{0},~H_{4}$  & $D_{4}$ \rule{0pt}{4mm} \\[2pt]\hline
$\Gamma_{2}~(A_{2})$ & $H_{z}^{\alpha}$          & $J_z,~T_{z}^{\alpha},~D_{z}^{1\alpha},D_{z}^{2\alpha}$ \rule{0pt}{4mm} \\[2pt]\hline
$\Gamma_{3}~(B_{1})$ & $O_{v},~H_{2}$            & $T_{xyz},~D_{2}$                  \rule{0pt}{4mm} \\[2pt]\hline 
$\Gamma_{4}~(B_{2})$ & $O_{xy},~H_{z}^{\beta}$ & $T_{z}^{\beta},~D_{z}^{\beta}$ \rule{0pt}{4mm} \\[2pt]\hline
$\Gamma_{5}~(E)$     &\begin{tabular}{c} $O_{zx},O_{yz}$ \rule{0pt}{4mm}\\ $H_{x}^{\alpha},~H_{y}^{\alpha},~H_{x}^{\beta},~H_{y}^{\beta}$\rule{0pt}{4mm} \end{tabular} &\begin{tabular}{c} $J_x,J_y,~T_{x}^{\alpha},T_{y}^{\alpha},~T_{x}^{\beta},T_{y}^{\beta}$ \rule{0pt}{4mm} \\ $D_{x}^{1\alpha},~D_{y}^{1\alpha},~D_{x}^{2\alpha},~D_{y}^{2\alpha},~D_{x}^{\beta},~D_{y}^{\beta}$ \rule{0pt}{4mm} \end{tabular} \rule{0pt}{0mm}\\\hline
\end{tabular}
}
\end{table} 

On the other hand, there also are differences between the two bands in the vicinity of $\varepsilon_{n\bm{k}}=\mu$ as seen in Fig.~\ref{Fig01}(b) and the corresponding Fermi surfaces~(FSs) in the tetragonal Brillouin zone~(BZ) for LaCoSi and CeCoSi are shown in Figs.~\ref{Fig01}(c) and (d), respectively. Three independent FSs are obtained for the band-index 31st, 33rd and 35th, where all bands have two-folded degeneracy due to the time-reversal~(TR) symmetry. The remarkable differences between the two bands can be seen especially in the 31st and 33rd FSs, where the 33rd FS forming elongated arms in the $\Gamma$-M direction in LaCoSi~[Fig.~\ref{Fig01}(c)] yields a square hole pocket centered at $\Gamma$ as the 31st FS in CeCoSi with the separation of the $\Gamma$-M arms~[Fig.~\ref{Fig01}(d)]. The reason for these differences is that the $c$ bands of the TB model have no complete $f$-electron component, whereas the DFT bands of LaCoSi inevitably contain the $f$-component due to the finite $c$-$f$ mixing. 
For the calculation of RKKY interactions as seen later, it is reasonable to utilize the present $c$ states corresponding to the localized $f$-electron limit without the $c$-$f$ mixing.

Next, we have extracted the one-body parameters for $f$ electrons from $\mathscr{H}_{f\textrm{-}f}$, such as the spin-orbit coupling~(SOC) constant for $\lambda_{\rm SOC}=94.2$~meV close to the typical value of Ce atom. The three KDs are obtained as $\Gamma_{6\pm}$~(0~meV), $\Gamma_{7a\pm}$~(9~meV) and $\Gamma_{7b\pm}$~(23~meV), respectively, which disagree the experimental CEF levels of $\Gamma_{7a\pm}$~(0~meV), $\Gamma_{7b\pm}$~(10~meV) and $\Gamma_{6\pm}$~(14~meV)\cite{Nikitin2020}. 
This contrasts with the results for CeB$_6$, where the ordering of the CEF levels in the experiment has been also obtained in the DFT model\cite{YH2019,YH2020}, possibly because the CEF splitting in the present system is smaller than in CeB$_6$. Here and hereafter, we adopt the experimental CEF levels in the following, since the $c$-$f$ mixing being crucial for the RKKY interaction is an one-body quantity between $f$ and $c$ electrons and should be almost unaffected by the CEF levels.

Here we investigate the single-site multipole susceptibility for the multipole $O_{\Gamma}$, 
\begin{align}
&\chi_{O_{_\Gamma}}^{(0)}=\sum_{m_1m_2m_3m_4}(O_{\Gamma})_{m_1m_2}\chi_{m_1m_2m_3m_4}^{(0)}(O_{\Gamma}^{\dagger})_{m_3m_4},
\end{align}
where $\chi_{m_1m_2m_3m_4}^{(0)}$ is the single-site susceptibility for 6 CEF states $\{m_{i}\}$, 
$|1\pm\rangle=\mp\sqrt{1-w^2}|\pm\frac{5}{2 }\rangle\pm w|\mp\frac{3}{2}\rangle~(\Gamma_{7a\pm})$, 
$|2\pm\rangle=w|\pm\frac{5}{2 }\rangle+\sqrt{1-w^2}|\mp\frac{3}{2}\rangle~(\Gamma_{7b\pm})$, 
$|3\pm\rangle=|\pm\frac{1}{2 }\rangle~(\Gamma_{6\pm})$ with $w=0.95$\cite{Nikitin2020}. 
6 CEF states have $6\times 6=36$ multipoles up to rank 5 classified into the irreducible representation~(IRR) of the Ce site under the point group $C_{4v}$ as shown in Table~\ref{table01}. 

\begin{figure}[t]
\centering
\includegraphics[width=7.5cm]{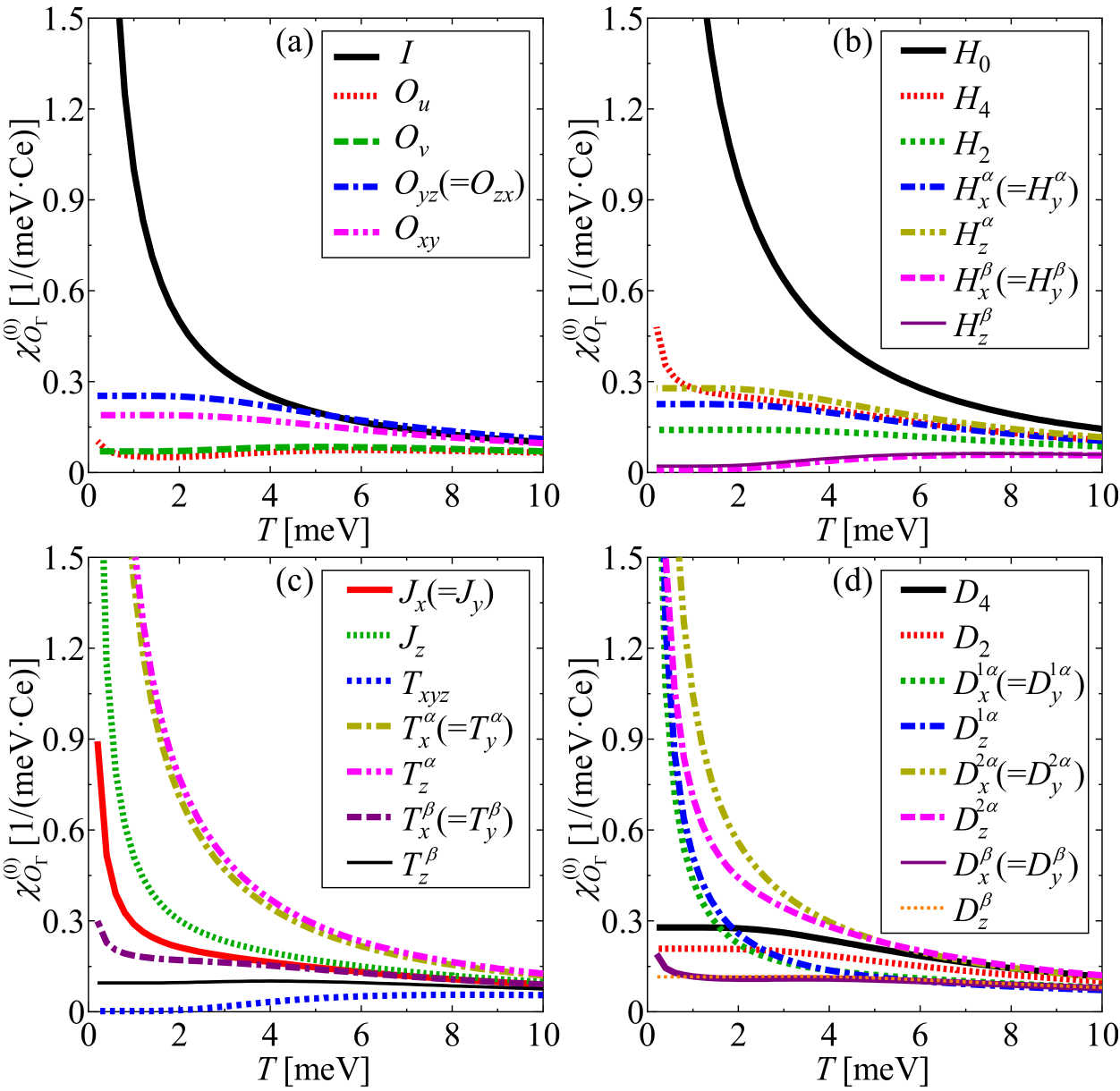}
\vspace{-2mm}
\caption{(Color online) 
$T$-dependence of $\chi_{O_{_{\Gamma}}}^{(0)}$ for (a) monopole and quadrupoles, (b) hexadecapoles, (c) dipoles and octupoles, and (d) triakontadipoles, where the $y$-component of $\chi_{O_{_{\Gamma}}}^{(0)}$~(not shown) is the same as the $x$-component due to the tetragonal symmetry, for example $\chi_{J_{y}}^{(0)}=\chi_{J_{x}}^{(0)}$.
}
\label{Fig02}
\end{figure}

Figure \ref{Fig02} shows the susceptibilities $\chi_{O_{_\Gamma}}^{(0)}$ for (a) monopole and quadrupoles, (b) hexadecapoles, (c) dipoles and octupoles, and (d) triakontadipoles as a function of $T$, where all multipoles $O_{\Gamma}$ are normalized as ${\rm Tr}[O_{\Gamma}^{2}]=6$ so as to compare all susceptibilities with the same footing. The susceptibilities only for the TR-even monopole $I$ and hexadecapole $H_0$~[Figs.~\ref{Fig02}(a) and (b)], the $z(x)$-components of TR-odd multipoles $J_{z}(J_{x})$, $T_{z}^{\alpha}(T_{x}^{\alpha})$, $D_{z}^{1\alpha}(D_{x}^{1\alpha})$, $D_{z}^{2\alpha}(D_{x}^{2\alpha})$~[Figs.~\ref{Fig02}(c) and (d)] show the Curie-like behavior in proportion to $1/T$ at low temperatures, since they have finite matrix elements between the lowest KD. Other susceptibilities such as $O_{yz/zx},O_{xy}$ and $H_{z}^{\beta},H_{x}^{\beta}$~[Figs.~\ref{Fig02}(a) and (b)] increase slightly with decreasing $T$ and then saturate to constant values, since they only have finite matrix elements between the lowest KD and the excited KDs, corresponding to Van-Vleck term. As a consequence, active multipoles at low temperatures are limited to those belonging to the same IRR under $C_{4v}$: 
TR-even $\Gamma_{1}$ multipoles $(I,~H_0,~O_u,H_4)$, and TR-odd $\Gamma_{2}$ and $\Gamma_{5}$ multipoles, $(J_{z},~T_{z}^{\alpha},~D_{z}^{1\alpha,2\alpha})$ and $(J_{x/y},~T_{x/y}^{\alpha},~D_{x/y}^{1\alpha,2\alpha})$, respectively.

\begin{figure*}[t]
\centering
\includegraphics[width=17.0cm]{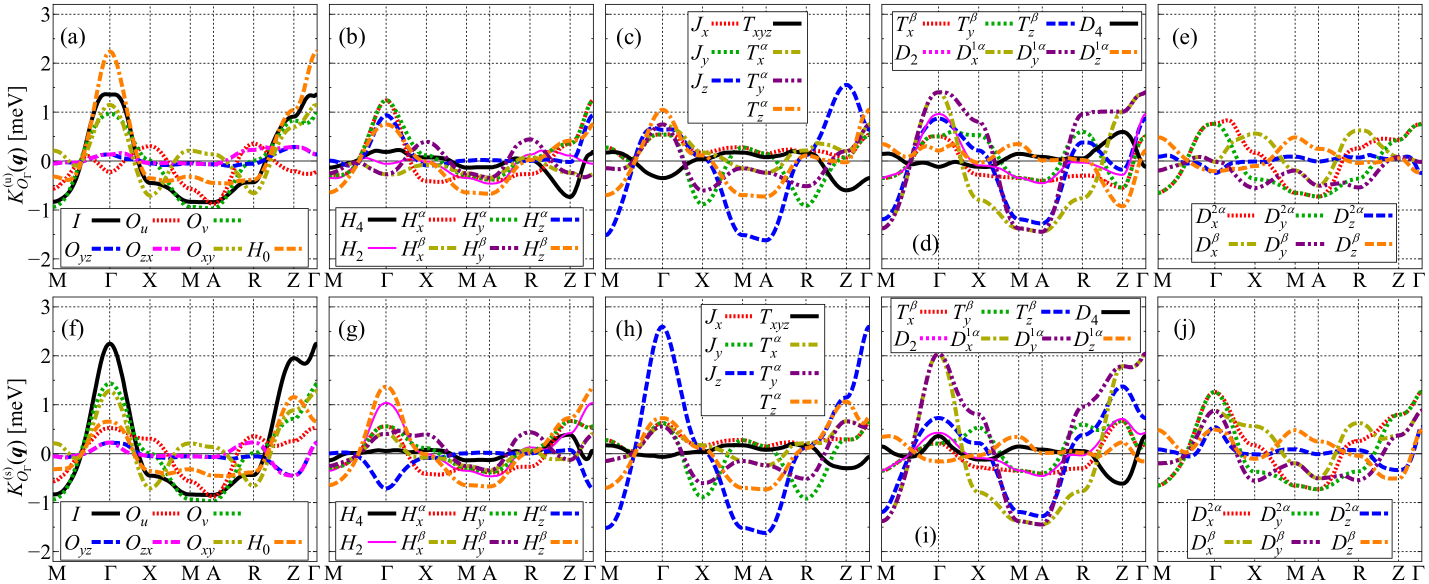}
\vspace{-2mm}
\caption{(Color online) 
$\bm{q}$-dependence of the uniform interaction $K_{O_{_\Gamma}}^{({\rm u})}(\bm{q})$ for (a),(b) TR-even and (c)-(e) TR-odd multipoles 
and the staggered interaction $K_{O_{_\Gamma}}^{({\rm s})}(\bm{q})$ for (f),(g) TR-even and (h)-(j) TR-odd multipoles along the high symmetry line in the BZ. 
}
\label{Fig03}
\end{figure*}

Then we calculate the RKKY interactions between 12 $f$ states/uc~(6 CEF states $m$ and 2 sites $\alpha={\rm Ce1},~{\rm Ce2}$) by the fourth-order perturbation w. r. t. $\mathscr{H}_{c-f}$\cite{YH2019,YH2020} which is given by,
\begin{align}
&\!\!K_{m_1m_2m_3m_4}^{\alpha,\beta}(\bm{q})\!=\!\frac{4}{\Delta_{0}^2}
\frac{1}{N}\!\sum_{\bm{k}nn'}\!
\mathcal{M}^{n\bm{k}}_{\beta m_3,\alpha m_1}
\mathcal{M}^{n\bm{k}+\bm{q}}_{\alpha m_2,\beta m_4}
\frac{f_{n'\bm{k}+\bm{q}}-f_{n\bm{k}}}
{\varepsilon_{n\bm{k}}-\varepsilon_{n'\bm{k}+\bm{q}}},
\label{eq:Kq}
\end{align}
where $\mathcal{M}^{n\bm{k}}_{\alpha m,\beta m'}=\sum_{\ell\ell'}V_{\bm{k}\ell\alpha m}^{*}V_{\bm{k}\ell'\beta m'}^{}U_{n\bm{k}\ell}^{c*}U_{n\bm{k}\ell'}^{c}$ and $V_{\bm{k}\ell\alpha m}$ is the one-body matrix element of $\mathscr{H}_{c\textrm{-}f}$ between $c~(\ell)$ and $f~(\alpha m)$ states and $U_{n\bm{k}\ell}^{c}$ is the eigenvector of $\mathscr{H}_{c\textrm{-}c}$ for $\ell$. $f_{n\bm{k}}$ is the Fermi distribution function $f_{n\bm{k}}=[\exp\{(\varepsilon_{n\bm{k}}-\mu)/\tilde{T}\}+1]^{-1}$, where we fix the temperature used for Eq.~(\ref{eq:Kq}) as $\tilde{T}=10~{\rm meV}$, since the RKKY interaction seems less temperature dependent in the present nesting-free FSs. By using Eq.~(\ref{eq:Kq}), we obtain the interactions for the uniform~(staggered) multipole $O_{\Gamma}^{({\rm u})}(O_{\Gamma}^{({\rm s})})=[O_{\Gamma}^{\rm Ce1}+(-)O_{\Gamma}^{\rm Ce2}]/\sqrt{2}$,
\begin{align}
&K_{O_{_\Gamma}}^{({\rm u,s})}(\bm{q})=\frac{1}{2}\sum_{\alpha\beta}\left[\left\{\delta_{\alpha\beta}(1\mp 1)\pm 1\right\}K_{O_{_\Gamma}}^{\alpha,\beta}(\bm{q})-\delta_{\alpha\beta}K_{O_{_\Gamma}}^{\alpha,\rm loc}\right],
\label{eq:KO}
\end{align}
where the double sign of up/down corresponds to u/s and $K_{O_{_\Gamma}}^{\alpha,\beta}(\bm{q})=(1/6)\sum_{\{m_i\}}(O^{\alpha}_{\Gamma})_{m_1m_2}K_{m_1m_2m_3m_4}^{\alpha,\beta}(\bm{q})(O_{\Gamma}^{\beta\dagger})_{m_3m_4}$ and $K_{O_{\Gamma}}^{\alpha,\rm loc}=(1/N)\sum_{\bm{q}}K_{O_{_\Gamma}}^{\alpha,\alpha}(\bm{q})$. In calculations, we use $\bm{k}$-mesh of $N=64^3$ and consider only $f^{0}$ intermediate excited state, since the contribution of $f^2$-process is found to be not primary for determining the multipole order from the study on CeB$_6$ based on the dynamical mean field theory with Hubbard I approximation\cite{Otsuki2022}. The excitation energy is set to $\Delta_{0}=2$~eV corresponding to the center of mass of the $f$ bands of LaCoSi shown in Fig.~\ref{Fig01}(a), which results in an energy denominator factor of 1 for the interaction $K(\bm{q})$~(see \S4 of SM\cite{SM}). 

The obtained $K_{O_{_\Gamma}}^{(\rm u)}(\bm{q})$ and $K_{O_{_\Gamma}}^{(\rm s)}(\bm{q})$ for all multipoles listed in Table~\ref{table01} are plotted along the high symmetry line in BZ shown in Figs.~\ref{Fig03}(a)-(e) and (f)-(j), respectively, where the positive (negative) value for a certain multipole $O_{\Gamma}^{({\rm u,s})}$ with $\bm{q}$ enhances (suppresses) the corresponding multipole ordering tendency. All $K_{O_{_\Gamma}}^{({\rm u,s})}(\bm{q})$ take the values in the range of $-2$~meV to $3$~meV and have nontrivial $\bm{q}$ and multipole dependence reflecting the $c$ band $\varepsilon_{n\bm{k}}$ and $c$-$f$ mixing $V_{\bm{k}\ell\alpha m}$ in the present DFT-based TB model with the symmetry of the tetragonal crystal, for example $O_{yz}$ and $O_{zx}$ have the same~(different) values in the $\Gamma$-M~($\Gamma$-X) direction. Many multipoles show a maximum at $\bm{q}=\bm{0}$: the largest component is the staggered dipole $J_{z}$~[Fig.~\ref{Fig03}(h)], the second largest is the uniform hexadecapole $H_{0}$~[Fig.~\ref{Fig03}(a)] and staggered monopole $I$~[Fig.~\ref{Fig03}(f)], the third is the staggered triakontadipole $D_{x,y}^{1\alpha}$~[Fig.~\ref{Fig03}(i)]. These values consist mainly of four nearest-neighbor couplings between Ce1-Ce2 and four next-nearest-neighbor couplings between Ce1-Ce1(Ce2-Ce2), whose signs and magnitudes are complicated by the type of multipoles. The detailed structure of these interactions in real space and their relation to FSs and $c$-$f$ mixing will be reported somewhere in the near future.

The RPA susceptibility matrix $\hat{\chi}(\bm{q})$ is given with the matrices of the single-site susceptibility $\hat{\chi}^{(0)}$ and the interaction $\hat{K}(\bm{q})$ as follows, 
\begin{align}
\hat{\chi}(\bm{q})=\hat{\chi}^{(0)}\left[\hat{1}-\hat{K}(\bm{q})\hat{\chi}^{(0)}\right]^{-1},
\label{eq:chiq}
\end{align}
where each matrix of Eq.~(\ref{eq:chiq}) has $2\times 6^2=72$ dimension corresponding to $i=(\alpha m_1m_2)$ of the onsite $f$ basis, leading to $[\hat{\chi}^{(0)}]_{i,j}=\delta_{\alpha\beta}\chi_{m_1m_2m_3m_4}^{(0)}$ and $[\hat{\chi}(\hat{K})]_{i,j}=\chi_{m_1m_2m_3m_4}^{\alpha,\beta}~(K_{m_1m_2m_3m_4}^{\alpha,\beta})$. By diagonalizing the matrix $\hat{K}(\bm{q})\hat{\chi}^{(0)}$ at each $\bm{q}$, we obtain the maximum~(first largest) eigenvalue $\alpha_{1}(\bm{q})$ which is enhanced towards a certain ordering instability with the ordered wavevector $\bm{q}=\bm{q}_{\rm max}$ and finally reaches $\alpha_{1}=1$ for the critical point of the multipole ordering phase transition. 

\begin{figure}[t]
\centering
\includegraphics[width=7.5cm]{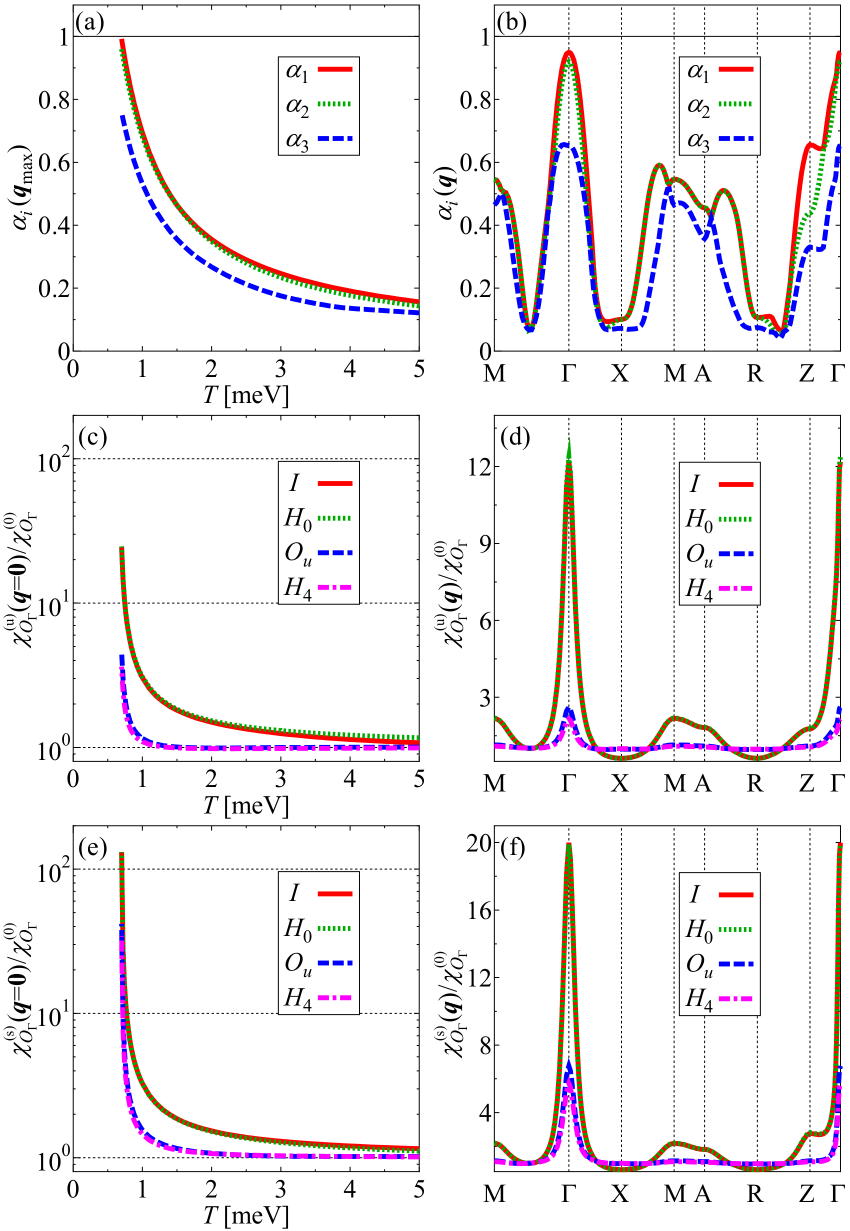}
\vspace{-2mm}
\caption{(Color online) 
(a) $T$-dependence of the $i$-th largest eigenvalue $\alpha_{i}(\bm{q}_{\rm max})$ of the matrix $\hat{K}(\bm{q})\hat{\chi}^{(0)}$ with $\bm{q}_{\rm max}=\bm{0}$. (c),(e) $T$-dependence of the enhancement rates of the $\bm{q}=\bm{0}$ RPA susceptibility for (c) uniform and (e) staggered multipoles $(I,~H_0,~O_u,~H_4)$ of the TR-even $\Gamma_{1}$. (b),(d),(f) The corresponding $\bm{q}$-dependences for $(\alpha_{1},T)=(0.95,~0.73~{\rm meV})$. 
}
\label{Fig04}
\end{figure}

Figure \ref{Fig04}(a) shows the $i$-th largest eigenvalue $\alpha_{i}(\bm{q}_{\rm max})$ up to 3rd values as a function of $T$. Each eigenvalue increases with decreasing $T$ and the maximum eigenvalue $\alpha_{1}=1$ is realized at $T=0.7$~meV with $\bm{q}_{\rm max}=\bm{0}$. Here, $T$-dependence of $\alpha_{i}(\bm{q})$ is due to the single-site susceptibility $\chi^{(0)}$ shown in Fig.~\ref{Fig02}, since the interaction $K(\bm{q})$ is independent of temperature as mentioned above. The interactions of the active multipoles at low temperatures show maximums at $\bm{q}=\bm{0}$ as shown in Fig.~\ref{Fig03}, so that the corresponding multipole fluctuations develop towards the multipole order at $\bm{q}=\bm{0}$. For explicit check, we plot the $\bm{q}$-dependence of $\alpha_{i}(\bm{q})$ shown in the inset of Fig.~\ref{Fig04}(b) for $(\alpha_{1},~T)=(0.95,~0.73~{\rm meV})$, where the clear peaks can be seen at $\bm{q}=\bm{0}$ for each $\alpha_{i}$ and the first and second eigenvalues are quite close to each other. 

In order to clarify such $\bm{q}=\bm{0}$ order obtained in Fig.~\ref{Fig04}(a), we examine the enhancement rate of the RPA susceptibility $\chi_{O_{_\Gamma}}^{({\rm u,s})}(\bm{q})/\chi_{O_{_\Gamma}}^{(0)}$ for all multipoles. As a results, we have found that the RPA susceptibilities for TR-even $\Gamma_{1}$ multipoles $(I,~H_0,~O_u,~H_4)$ dominate over the other multipoles. 

Figures~\ref{Fig04}(c)-(f) show the enhancement rates for the TR-even $\Gamma_1$ multipoles of (c) uniform and (e) staggered components as a function of $T$, where Figs.~\ref{Fig04}(d) and (f) show the $\bm{q}$-dependence of the corresponding enhancement rates for the same values of Fig.~\ref{Fig04}(b). With decreasing $T$, $\chi_{O_{_\Gamma}}^{({\rm u})}/\chi_{O_{_\Gamma}}^{(0)}$ of the monopole $I$ and hexadecapole $H_{0}$ increase rapidly with keeping the same value as shown in Fig.~\ref{Fig04}(c). A similar enhancement is also observed in $\chi_{O_{_\Gamma}}^{({\rm s})}/\chi_{O_{_\Gamma}}^{(0)}$ for $I$ and $H_{0}$ as shown in Fig.~\ref{Fig04}(e), but the staggered component is larger than the uniform one as compered to Fig.~\ref{Fig04}(c) and (d), where $\chi_{I}^{({\rm s})}/\chi_{I}^{(0)}\simeq\chi_{H_{0}}^{({\rm s})}/\chi_{H_{0}}^{(0)}\simeq 20$ shown in Fig.~\ref{Fig04}(f) is consistent with the enhancement rate from Eq.~(\ref{eq:chiq}) $1/(1-\alpha_{1})=20$ times, corresponding to the maximum eigenmode in Fig.~\ref{Fig04}(b). 
As a consequence, an antiferro multipole order mixed with $(I,~H_{0})$ is realized due to the $\bm{q}=\bm{0}$ staggered fluctuation as the maximum RPA mode, while the $\bm{q}=\bm{0}$ uniform fluctuation of $(I,~H_{0})$ inducing a phase separation instability is the second largest one which is not realized in actually. The fluctuations of $O_u$ and $H_{4}$ also increase with keeping slightly small value compared to the main components $I$ and $H_{0}$ as shown in Fig.~\ref{Fig04}(c)-(f), since the single-site susceptibility and interactions for different multipoles belonging to the same IRR~$\Gamma_1$ are also finite (not shown). We also investigate the rates for other multipoles except for the TR-even $\Gamma_{1}$ explicitly shown in \S5 of SM\cite{SM}, where the rates for the quadrupoles and hexadecapoles have a small $T$-dependence and are hardly enhanced, while that for TR-odd multipoles of $(J_{z},~T_{z}^{\alpha},~D_{z}^{1\alpha,2\alpha})~[\Gamma_{2}]$ and $(J_{x/y},~T_{x/y}^{\alpha},~D_{x/y}^{1\alpha,2\alpha})~[\Gamma_{5}]$ enhance two to three times. 

\begin{figure}[t]
\centering
\includegraphics[width=7.5cm]{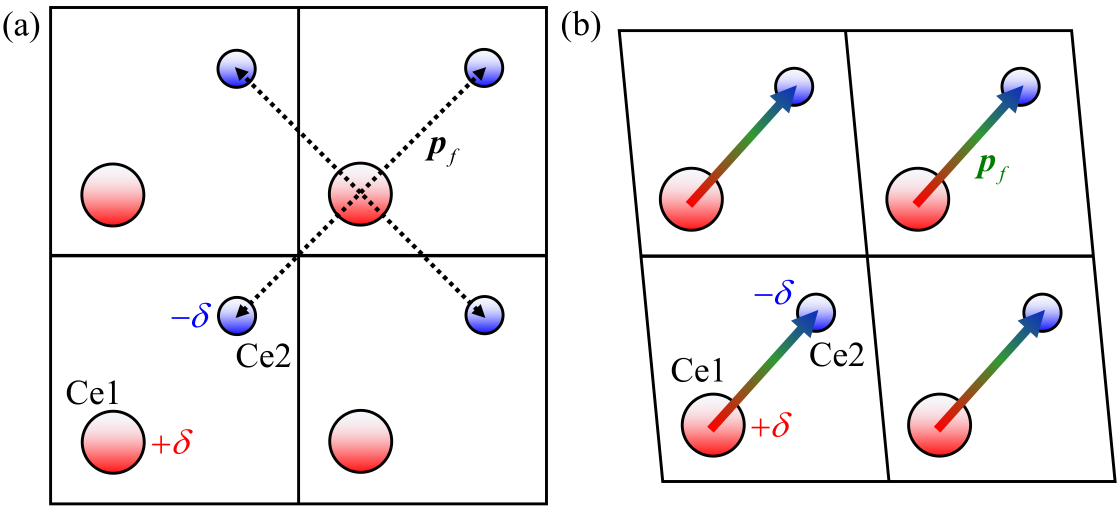}
\vspace{-2mm}
\caption{(Color online) 
Schematic pictures of the $f$-CDW order in the (a) tetragonal and (b) triclinic lattices, where $f$-electron electric dipoles $\bm{p}_{f}$ are dipicted by (a) dotted and (b) solid vectors, and large red (small blue) circles indicate an increase (decrease) in the number of $f$ electrons at Ce atoms.
}
\label{Fig05}
\vspace{+0.2cm}
\end{figure}

Here we discuss the relationship between the present results and previous studies and experiments. First, the obtained antiferro order of $(I,~H_{0})$ is consistent with the HO in the experiments in the sense of nonmagnetic\cite{Tanida2018,Tanida2019,Manago2021}. This $(I,~H_{0})$ order is also a $f$-electron charge density wave~($f$-CDW) with $\bm{q}=\bm{0}$, yielding the $f$-electron number $n_{f}$ imbalance on the two Ce atoms in uc, where $n_{f}=1$ on each Ce atom in $T>T_0$ splits into $n_{f}=1+(-)\delta$ on the Ce1~(Ce2) atom in $T<T_0$ as shown in Fig.~\ref{Fig05}(a). Similar antiferro charge ordering in $f$-electron systems has been discussed in studies of the filled skutterudite compounds PrRu$_4$P$_{12}$\cite{Takimoto2006} and SmRu$_4$P$_{12}$\cite{Shiina2013}, where $f$-electron charge ordering is driven by the perfect nesting of $c$-electron FS and realized in both metal-insulator transition. This is contrast to the present $f$-CDW driven by the RKKY interactions derived from the DFT bandstructure without the nesting, where $\bm{q}=\bm{0}$ charge ordering is realized on the two Ce atoms in uc and the $c$-electron system remains metal. 

The deviation of $n_f$ from 1 in the present $f$-CDW is considered to be small~($\delta\ll 1$), and some of them are expected to transfer to the $c$ electron system, i. e., the $f$ electrons should become somewhat itinerant for $T<T_0$, which is supported by the optical conductivity experiment\cite{Kimura2023}. In addition, the present $f$-CDW is coupled to the hexadecapole $H_0$, yielding to a shift in the CEF levels with an opposite splitting at the two Ce atoms, which may be related to the mysterious excitation observed in neutron scattering\cite{Nikitin2020}. Moreover, the contributions of $I$ and $H_0$ to the $f$-CDW would vary gradually with pressure, which could also explain the $T_0$ dip observed in electrical resistivity under high pressure\cite{Lengyel2013} by accompanying the CDW of the $c$-electrons, where the $f$-electrons are more itinerant. 

Finally, under the present $f$-CDW order, four equivalent electric dipoles $\bm{p}_{f}$ connecting Ce1-Ce2 are realized as shown in Fig.~\ref{Fig05}(a), and when the triclinic distortion occurs one of them becomes inequivalent and a single electric dipole with a vector connecting the intracell Ce atoms in uc is chosen as shown in Fig.~\ref{Fig05}(b). The triclinic structural phase transition in the experiments\cite{Matsumura2022,Manago2023} may be regarded as a consequence of the energy stabilizing of such the induced electric dipole alignment. Alternatively, it may be necessary to introduce the electron-lattice interactions into the present model in order to describe the structural transition and one of the important issues to be addressed in the future. A unified understanding of such the structural phase transition at ambient pressure and anomalies at high pressure is an important issue that can be positioned as an extension of this study.

In summary, we have studied the RKKY interactions between $f$ electrons based on the realistic Wannier model derived from the ab initio calculations on CeCoSi. The RPA analysis reveals that the leading order is a staggered $\bm{q}=\bm{0}$ monopole order coupled with hexadecapole $H_0$, which corresponds to a $f$-CDW with a staggered shift in the crystal field splitting and can explain the some experiments in the HO.

\begin{acknowledgment}
We would like to thank H. Tanida for valuable comments and fruitful discussions. This work was supported by JSPS KAKENHI Grant Numbers JP24K06943 and JP21H01031.
\end{acknowledgment}

\bibliography{CeCoSi-RKKY-cmat}



\clearpage
\onecolumn
\begin{center}
{\normalsize
{\bf
Supplemental material to ``RKKY Interactions and Multipole Order in Ab initio Wannier Model of CeCoSi''
}\\
Takemi Yamada, Yuki Yanagi and Keisuke Mitsumoto\\
\medskip
\textit{Liberal Arts and Sciences, Toyama Prefectural University, Imizu, Toyama 939-0398, Japan}\\
\medskip
}
\end{center}

\vspace{-2mm}

\normalsize{
This supplemental material contains: 
1. DFT calculation conditions for WIEN2k, 
2. TB model construction for Wannier90, 
3. Multipole operator and single-site susceptibility, 
4. Derivation of RKKY interactions and 
5. Enhancement rates of the RPA susceptibilities. 

\section{DFT calculation for WIEN2k}
Here we provide the details of the DFT calculation based on WIEN2k~\cite{Blaha2020}, which is the all-electron first-principles code, where the basis functions are expanded by the relativistic full-potential augmented plane wave~(FLAPW) and/or APW + local orbitals depending on the core and valence states of each atom in uc. We employ the generalized gradient approximation for the exchange-correlation potential of the PBE-GGA potential~\cite{PBE-GGA1996}. The SOC is fully included within the second variation approximation. 

In self-consistent calculation, we use the experimental lattice parameters\cite{Tanida2019} in the space group $P4/nmm$~(\#129), 
the lattice constants $a=b=4.057~{\rm \AA}$ and $c=6.987~{\rm \AA}$, the internal fractional coordinates $(x,y,z)=(0.25,0.25,0.6781)$ for Ce, $(0.75,0.25,0)$ for Co and $(0.25,0.25,0.183)$ for Si, and the calculation conditions: $\bm{k}$-mesh of $20^2\times 23$ corresponding to 2772 $\bm{k}$-points in the irreducible part of BZ and the muffin-tin radius $R_{\rm MT}^{\rm Ce,Co(Si)}=2.5~(1.91)$ bohr. The plane wave cutoff is $R_{\rm MT}K_{\rm max}=9$ and the energy and charge convergence conditions are set to 10$^{-6}$~Ry and 10$^{-5}$~e, respectively.

\section{TB model construction by Wannier90}
We use Wannier90\cite{w90-Pizzi2020} to construct a TB model of CeCoSi with localized Wannier orbitals, which are given by the inverse Fourier transform of the DFT Bloch states and thus can fully reproduce the DFT bands within a certain energy window. 
In particular, well localized Wannier orbitals at each atomic center can be treated as atomic orbitals, allowing for more detailed microscopic analysis than the DFT band calculation itself and incorporating many-body interactions not included in the DFT calculation, such as the RKKY interaction calculated in this study. 

In order to simultaneously achieve the main valence bandstructure and Wannier function localization at a high level, we construct a TB model with 84 localized Wannier orbitals, which consists of Ce-$4f,5d~(7+5=12)$, Co-$3d~(5)$, and Si-$3p,4s~(3+1=4)$ for a total of $12+5+4=21$ orbitals with spin~(2) degrees of freedom and the sub-lattice~(2) degrees of freedom by 2CeCoSi/uc, resulting in $21\times 2\times 2=84$ orbitals. The inner window~(dis-frozen energy window) corresponding to the region reproducing the bandstructure is set in the energy range from $-10~{\rm eV}$ to $+4~{\rm eV}$ and the target band for adjusting the outer window is set to $84+180=264$, where the localization of the Wannier function improves as the target band increases, though in this study the qualitative results are almost unchanged by the case of about $84+20=104$ bands. The $\bm{k}$-mesh is set to $8^2\times 6=384$ corresponding to $567$ real-space hopping vectors and the disentangle convergence condition is set to 10$^{-7}~\AA^2$.

The obtained TB model is given by the following form, 
\begin{align}
&\mathscr{H}_{\rm TB}=\mathscr{H}_{c\textrm{-}c}+\mathscr{H}_{f\textrm{-}f}+\mathscr{H}_{c\textrm{-}f},\label{eq:HTB}\\
&\quad\mathscr{H}_{c\textrm{-}c}=\sum_{\bm{i},\bm{\delta}}\sum_{\ell\ell'}h_{\bm{i}\ell\ell'}^{cc}(\bm{\delta})c_{\bm{i}\ell}^{\dagger}c_{\bm{i}+\bm{\delta}\ell'}^{},\\
&\quad\mathscr{H}_{f\textrm{-}f}=\sum_{\bm{i},\bm{\delta}}\sum_{mm'}h_{\bm{i}mm'}^{ff}(\bm{\delta})f_{\bm{i}m}^{\dagger}f_{\bm{i}+\bm{\delta}m'}^{},\\
&\quad\mathscr{H}_{c\textrm{-}f}=\sum_{\bm{i},\bm{\delta}}\sum_{\ell m}\left(V_{\bm{i}\ell m}(\bm{\delta})c_{\bm{i}\ell}^{\dagger}f_{\bm{i}+\bm{\delta}m}^{}+h.c.\right),
\end{align}
where $c_{\bm{i}\ell}^{\dagger}~(f_{\bm{i}m}^{\dagger})$ is a creation operator of $c~(f)$ electron in uc at $\bm{R_i}$ with a state $\ell~(m)$ being 56 $d,p,s$ real functions per spin on each atom~(28 CEF eigenstates with total angular momentum $j=5/2,7/2$ on each Ce atom). 
The $c$-$c$~($f$-$f$) matrix element of $h_{\bm{i}\ell\ell'}^{cc}~(h_{\bm{i}mm'}^{ff})$ includes the $c~(f)$ energy levels, SOC couplings, CEF splittings and $c$-$c$~($f$-$f$) hopping integrals with the real-space hopping vector $\bm{\delta}$. 
$V_{\bm{i}\ell m}(\bm{\delta})$ is the $c$-$f$ mixing element, which is finite in both the onsite and intersite components due to the lack of inversion symmetry at each atom. 

The wavevector $\bm{k}$-representation of $\mathscr{H}_{\rm TB}$ is given by, 
\begin{align}
\mathscr{H}_{\rm TB}
&=\sum_{\bm{k}}\sum_{\ell\ell'}h_{\ell\ell'}^{cc}(\bm{k})c_{\bm{k}\ell}^{\dagger}c_{\bm{k}\ell'}^{}
+\sum_{\bm{k}}\sum_{mm'}h_{mm'}^{ff}(\bm{k})f_{\bm{k}m}^{\dagger}f_{\bm{k}m'}^{}
+\sum_{\bm{k}}\sum_{\ell m}\left(V_{\bm{k}\ell m}c_{\bm{k}\ell}^{\dagger}f_{\bm{k}m}^{}+h.c.\right),
\label{eq:HTB-k}
\end{align}
and $\mathscr{H}_{\rm TB}$ can be diagonalized in $\bm{k}$-space and transformed into the eigenstates with band-index $n$ as follows, 
\begin{align}
\mathscr{H}_{\rm TB}&=\sum_{n\bm{k}}E_{n\bm{k}}a_{n\bm{k}}^{\dagger}a_{n\bm{k}}^{},
\quad a_{n\bm{k}}=\sum_{m}U_{n\bm{k}m}f_{\bm{k}m}+\sum_{\ell}U_{n\bm{k}\ell}c_{\bm{k}\ell}, 
\end{align}
where the eigenenergy $E_{n\bm{k}}$ is the full TB bands reproducing the DFT bands of CeCoSi and 
an annihilation operator $a_{n\bm{k}}$ can be written by the the eigenvector components $U_{n\bm{k}m}$ and $U_{n\bm{k}\ell}$. 
Here we note that $U_{n\bm{k}\ell}$ is the eigenvector of $E_{n\bm{k}}$ and different from $U_{n\bm{k}\ell}^{c}$ which is the eigenvector of $\varepsilon_{n\bm{k}}$ in the main text. 

The obtained $E_{n\bm{k}}$ describe the itinerant $f$-electron bands and disagree with the electronic structure suggested by the ARPES experiment\cite{Kimura2021}, while as mentioned in the main text, the diagonalized $c$-electron bands of only $\mathscr{H}_{c\textrm{-}c}$ shows relatively good correspondence with the ARPES\cite{Kimura2021}, which justifies starting from the localized $f$-electron limit where $f$ electrons are well localized at each Ce.

\section{Multipole operator and single-site susceptibility}
Here we briefly address the multipole operator and single-site susceptibility described in the main text. 
As multipoles in this study, we treat only the traditional local multipoles given by $6\times 6$ matrices for a Ce-$4f$ one-electron basis $|j=5/2,m=-j,\cdots,j-1,j\rangle$, i.e., electric multipoles and magnetic multipoles\cite{Kusunose2008,Hayami2018}. 
In this case, we can define $6\times 6=36$ independent operators described by a spherical tensor operator $T_{q}^{(k)}$ with the same symmetry as the spherical harmonic function on isolated ions in a vacuum, where $k$ is rank and $q=-k,-k+1,\cdots,+k$. 
In crystals, we can also define the tesseral tensor operator $T_{kq}^{\rm (c,s)}$ given by a linear combination of $T_{q}^{(k)}$ as follows, 
\begin{align}
T_{k0}^{\rm (c)}&=T_{0}^{(k)},\\
T_{kq}^{\rm (c)}&=(-1)^{q}\frac{1}{\sqrt{2}}\left\{\left(T_{q}^{(k)}\right)^{\dagger}+T_{q}^{(k)}\right\}
=\begin{cases}
\cfrac{1}{\sqrt{2}}\left(T_{-q}^{(k)}-T_{q}^{(k)}\right) & (q={\rm odd}) \\
\cfrac{1}{\sqrt{2}}\left(T_{-q}^{(k)}+T_{q}^{(k)}\right) & (q={\rm even}) \\
\end{cases},\\
T_{kq}^{\rm (s)}&=(-1)^{q}\frac{i}{\sqrt{2}}\left\{\left(T_{q}^{(k)}\right)^{\dagger}-T_{q}^{(k)}\right\}
=\begin{cases}
\cfrac{i}{\sqrt{2}}\left(T_{-q}^{(k)}+T_{q}^{(k)}\right) & (q={\rm odd}) \\
\cfrac{i}{\sqrt{2}}\left(T_{-q}^{(k)}-T_{q}^{(k)}\right) & (q={\rm even}) \\
\end{cases},
\end{align}
where $\left(T_{q}^{(k)}\right)^{\dagger}=T_{-q}^{(k)}$ and the linear combination of the even~(odd) rank multipoles corresponds to the electric~(magnetic) multipoles. 

Table~\ref{table-S1} summarizes the classification of multipole operators in CeCoSi under the IRR of the point group $C_{4v}$ at each Ce site and their explicit representation by tesseral tensor operators, where the multipole operators are written in the same notation as the study on the similar tetragonal compound URu$_2$Si$_2$, such as monopole (dipole), quadrupole (octupole) and hexadecapole (triakontadipole) are denoted by $I~(J)$, $O~(T)$, and $H~(D)$, respectively\cite{Ikeda2012}. 

\begin{table*}[t]
\caption{
List of all multipole operators $O_{\Gamma}$ up to rank 5 and their IRRs in the point groups $O$ and $C_{4v}$ with $(x,y,z)$- and $T_{kq}^{\rm (c,s)}$-notations, where the double sign of IRR $\Gamma_{\gamma}^{\pm}$ corresponds to the TR-even/TR-odd symmetry.
}\label{table-S1}
\vspace{2mm}
\centering
\scalebox{0.75}{
\begin{tabular}{cccccc}
\hline
rank &$O$ &$C_{4v}$ &$O_{\Gamma}$ &$(x,y,z)$-notation &$T_{kq}^{\rm (c,s)}$-notation \\
\hline\hline
0&$\Gamma_{1}^{+}(A_{1}^{+})$ &$\Gamma_{1}^{+}(A^{+})$     &$I$                  &$1$ &$T_{00}^{\rm (c)}$ \rule{0pt}{6mm} \\\hline
1&$\Gamma_{4}^{-}(T_{1}^{-})$ &$\Gamma_{5}^{-}(E^{-})$     &$J_{x}$              &$x$ &$T_{11}^{\rm (c)}$ \rule{0pt}{6mm}\\
 &                                     &$\Gamma_{5}^{-}(E^{-})$     &$J_{y}$             &$y$ &$T_{11}^{\rm (s)}$ \rule{0pt}{6mm}\\
 &                                     &$\Gamma_{2}^{-}(A_{2}^{-})$ &$J_{z}$             &$z$ &$T_{10}^{\rm (c)}$  \rule{0pt}{6mm}\\\hline
2&$\Gamma_{3}^{+}(E^{+})$     &$\Gamma_{1}^{+}(A_{1}^{+})$  &$O_{u}$            &$\tfrac{1}{2}(3z^2-r^2)$  &$T_{20}^{\rm (c)}$  \rule{0pt}{6mm}\\
 &                                    &$\Gamma_{3}^{+}(B_{1}^{+})$  &$O_{v}$            &$\tfrac{\sqrt{3}}{2}(x^2-y^2)$ &$T_{22}^{\rm (c)}$ \rule{0pt}{6mm}\\\hdashline
 &$\Gamma_{5}^{+}(T_{2}^{+})$ &$\Gamma_{5}^{+}(E^{+})$      &$O_{yz}$           &$\sqrt{3}yz$ &$T_{21}^{\rm (s)}$ \rule{0pt}{6mm}\\
 &                                    &$\Gamma_{5}^{+}(E^{+})$      &$O_{zx}$           &$\sqrt{3}zx$ &$T_{21}^{\rm (c)}$ \rule{0pt}{6mm}\\
 &                                    &$\Gamma_{4}^{+}(B_{2}^{+})$  &$O_{xy}$           &$\sqrt{3}xy$ &$T_{22}^{\rm (s)}$ \rule{0pt}{6mm}\\\hline
3&$\Gamma_{2}^{-}(A_{2}^{-})$ &$\Gamma_{3}^{-}(A_{2}^{-})$ &$T_{xyz}$          &$\sqrt{15}xyz$ &$T_{32}^{\rm (s)}$ \rule{0pt}{6mm}\\\hdashline
 &$\Gamma_{4}^{-}(T_{1}^{-})$ &$\Gamma_{5}^{-}(E^{-})$      &$T_{x}^{\alpha}$  &$\tfrac{1}{2}x(5x^2-3r^2)$ &$\sqrt{\tfrac{5}{8}}T_{33}^{\rm (c)}-\sqrt{\tfrac{3}{8}}T_{31}^{\rm (c)}$ \rule{0pt}{6mm}\\
 &                                    &$\Gamma_{5}^{-}(E^{-})$      &$T_{y}^{\alpha}$  &$\tfrac{1}{2}y(5y^2-3r^2)$ &$-\sqrt{\tfrac{5}{8}}T_{33}^{\rm (s)}-\sqrt{\tfrac{3}{8}}T_{31}^{\rm (s)}$ \rule{0pt}{6mm}\\
 &                                    &$\Gamma_{2}^{-}(A_{2}^{-})$  &$T_{z}^{\alpha}$  &$\tfrac{1}{2}z(5z^2-3r^2)$ &$T_{30}$ \rule{0pt}{6mm}\\\hdashline
 &$\Gamma_{5}^{-}(T_{2}^{-})$ &$\Gamma_{5}^{-}(E^{-})$      &$T_{x}^{\beta}$   &$\tfrac{\sqrt{15}}{2}x(y^2-z^2)$ &$-\sqrt{\tfrac{3}{8}}T_{33}^{\rm (c)}-\sqrt{\tfrac{5}{8}}T_{31}^{\rm (c)}$ \rule{0pt}{6mm}\\
 &                                    &$\Gamma_{5}^{-}(E^{-})$      &$T_{y}^{\beta}$   &$\tfrac{\sqrt{15}}{2}y(z^2-x^2)$ &$-\sqrt{\tfrac{3}{8}}T_{33}^{\rm (s)}+\sqrt{\tfrac{5}{8}}T_{31}^{\rm (s)}$ \rule{0pt}{6mm}\\
 &                                    &$\Gamma_{4}^{-}(B_{2}^{-})$ &$T_{z}^{\beta}$   &$\tfrac{\sqrt{15}}{2}z(x^2-y^2)$ &$T_{32}^{\rm (c)}$ \rule{0pt}{6mm}\\\hline
4&$\Gamma_{1}^{+}(A_{1}^{+})$ &$\Gamma_{1}^{+}(A_{1}^{+})$ &$H_{0}$            &$\tfrac{5\sqrt{21}}{12}(x^4+y^4+z^4-\tfrac{3}{5}r^4)$ &$\sqrt{\tfrac{5}{12}}T_{44}^{\rm (c)}+\sqrt{\tfrac{7}{12}}T_{40}^{\rm (c)}$ \rule{0pt}{6mm}\\\hdashline
 &$\Gamma_{3}^{+}(E^{+})$      &$\Gamma_{1}^{+}(A_{1}^{+})$ &$H_{4}$            &$\tfrac{7\sqrt{15}}{12}[2z^4-x^4-y^4-\tfrac{6}{7}r^2(3z^2-r^2)]$ &$-\sqrt{\tfrac{7}{12}}T_{44}^{\rm (c)}+\sqrt{\tfrac{5}{12}}T_{40}^{\rm (c)}$ \rule{0pt}{6mm}\\
 &                                    &$\Gamma_{3}^{+}(B_{1}^{+})$ &$H_{2}$            &$\tfrac{7\sqrt{5}}{4}[x^4-y^4-\tfrac{6}{7}r^2(x^2-y^2)]$ &$-T_{42}^{\rm (c)}$ \rule{0pt}{6mm}\\\hdashline
 &$\Gamma_{4}^{+}(T_{1}^{+})$ &$\Gamma_{5}^{+}(E^{+})$      &$H_{x}^{\alpha}$ &$\tfrac{\sqrt{35}}{2}yz(y^2-z^2)$ &$-\sqrt{\tfrac{1}{8}}T_{43}^{\rm (s)}-\sqrt{\tfrac{7}{8}}T_{41}^{\rm (s)}$ \rule{0pt}{6mm}\\
 &                                    &$\Gamma_{5}^{+}(E^{+})$      &$H_{y}^{\alpha}$ &$\tfrac{\sqrt{35}}{2}zx(z^2-x^2)$ &$-\sqrt{\tfrac{1}{8}}T_{43}^{\rm (c)}+\sqrt{\tfrac{7}{8}}T_{41}^{\rm (c)}$ \rule{0pt}{6mm}\\
 &                                    &$\Gamma_{2}^{+}(A_{2}^{+})$  &$H_{z}^{\alpha}$ &$\tfrac{\sqrt{35}}{2}xy(x^2-y^2)$ &$T_{44}^{\rm (s)}$ \rule{0pt}{6mm}\\\hdashline
 &$\Gamma_{5}^{+}(T_{2}^{+})$ &$\Gamma_{5}^{+}(E^{+})$      &$H_{x}^{\beta}$   &$\tfrac{\sqrt{5}}{2}yz(7x^2-r^2)$ &$\sqrt{\tfrac{7}{8}}T_{43}^{\rm (s)}-\sqrt{\tfrac{1}{8}}T_{41}^{\rm (s)}$ \rule{0pt}{6mm}\\
 &                                    &$\Gamma_{5}^{+}(E^{+})$      &$H_{y}^{\beta}$   &$\tfrac{\sqrt{5}}{2}zx(7y^2-r^2)$ &$-\sqrt{\tfrac{7}{8}}T_{43}^{\rm (c)}-\sqrt{\tfrac{1}{8}}T_{41}^{\rm (c)}$ \rule{0pt}{6mm}\\
 &                                    &$\Gamma_{4}^{+}(B_{2}^{+})$ &$H_{z}^{\beta}$   &$\tfrac{\sqrt{5}}{2}xy(7z^2-r^2)$ &$T_{42}^{\rm (s)}$ \rule{0pt}{6mm}\\\hline
5&$\Gamma_{3}^{-}(E^{-})$     &$\Gamma_{1}^{-}(A_{1}^{-})$ &$D_{4}$             &$\tfrac{3\sqrt{35}}{2}xyz(x^2-y^2)$    &$T_{54}^{\rm (s)}$  \rule{0pt}{6mm}\\
 &                                    &$\Gamma_{3}^{-}(B_{1}^{-})$ &$D_{2}$             &$-\tfrac{\sqrt{105}}{2}xyz(3z^2-r^2)$ &$-T_{52}^{\rm (s)}$  \rule{0pt}{6mm}\\\hdashline
 &$\Gamma_{4}^{-}(T_{1}^{-})$ &$\Gamma_{5}^{-}(E^{-})$      &$D_{x}^{1\alpha}$ &$\tfrac{x}{8}\left[8x^4-40x^2(y^2+z^2)+15(y^2+z^2)^2\right]$ &$\tfrac{1}{8\sqrt{2}}(3\sqrt{7}T_{55}^{\rm (c)}-\sqrt{35}T_{53}^{\rm (c)}+\sqrt{30}T_{51}^{\rm (c)})$ \rule{0pt}{6mm}\\
 &                                    &$\Gamma_{5}^{-}(E^{-})$      &$D_{y}^{1\alpha}$ &$\tfrac{y}{8}\left[8y^4-40y^2(z^2+x^2)+15(z^2+x^2)^2\right]$ &$\tfrac{1}{8\sqrt{2}}(3\sqrt{7}T_{55}^{\rm (s)}+\sqrt{35}T_{53}^{\rm (s)}+\sqrt{30}T_{51}^{\rm (s)})$  \rule{0pt}{6mm}\\
 &                                    &$\Gamma_{2}^{-}(A_{2}^{-})$ &$D_{z}^{1\alpha}$ &$\tfrac{z}{8}\left[8z^4-40z^2(x^2+y^2)+15(x^2+y^2)^2\right]$ &$T_{50}$  \rule{0pt}{6mm}\\\hdashline
 &$\Gamma_{4}^{-}(T_{1}^{-})$ &$\Gamma_{5}^{-}(E^{-})$     &$D_{x}^{2\alpha}$ &$\tfrac{3\sqrt{35}}{2}x\left[y^4+z^4-\tfrac{3}{4}(y^2+z^2)^2\right]$ &$\tfrac{1}{16}(\sqrt{10}T_{55}^{\rm (c)}+9\sqrt{2}T_{53}^{\rm (c)}+2\sqrt{21}T_{51}^{\rm (c)})$ \rule{0pt}{6mm}\\
 &                                    &$\Gamma_{5}^{-}(E^{-})$     &$D_{y}^{2\alpha}$ &$\tfrac{3\sqrt{35}}{2}y\left[z^4+x^4-\tfrac{3}{4}(z^2+x^2)^2\right]$ &$\tfrac{1}{16}(\sqrt{10}T_{55}^{\rm (s)}-9\sqrt{2}T_{53}^{\rm (s)}+2\sqrt{21}T_{51}^{\rm (s)})$  \rule{0pt}{6mm}\\
 &                                    &$\Gamma_{2}^{-}(A_{2}^{-})$ &$D_{z}^{2\alpha}$ &$\tfrac{3\sqrt{35}}{2}z\left[x^4+y^4-\tfrac{3}{4}(x^2+y^2)^2\right]$ &$T_{54}^{\rm (c)}$ \rule{0pt}{6mm}\\\hdashline
 &$\Gamma_{5}^{-}(T_{2}^{-})$ &$\Gamma_{5}^{-}(E^{-})$     &$D_{x}^{\beta}$    &$\tfrac{\sqrt{105}}{4}x(y^2-z^2)(3x^2-r^2)$ &$\tfrac{1}{4\sqrt{2}}(-\sqrt{15}T_{55}^{\rm (c)}-\sqrt{3}T_{53}^{\rm (c)}+\sqrt{14}T_{51}^{\rm (c)})$ \rule{0pt}{6mm}\\
 &                                    &$\Gamma_{5}^{-}(E^{-})$     &$D_{y}^{\beta}$    &$\tfrac{\sqrt{105}}{4}y(z^2-x^2)(3y^2-r^2)$ &$\tfrac{1}{4\sqrt{2}}(\sqrt{15}T_{55}^{\rm (s)}-\sqrt{3}T_{53}^{\rm (s)}-\sqrt{14}T_{51}^{\rm (s)})$  \rule{0pt}{6mm}\\
 &                                    &$\Gamma_{4}^{-}(B_{2}^{-})$ &$D_{z}^{\beta}$   &$\tfrac{\sqrt{105}}{4}z(x^2-y^2)(3z^2-r^2)$ &$T_{52}^{\rm (c)}$  \rule{0pt}{6mm}\\
\hline
\end{tabular}
}
\end{table*}

\begin{figure*}[t]
\centering
\includegraphics[width=12.0cm]{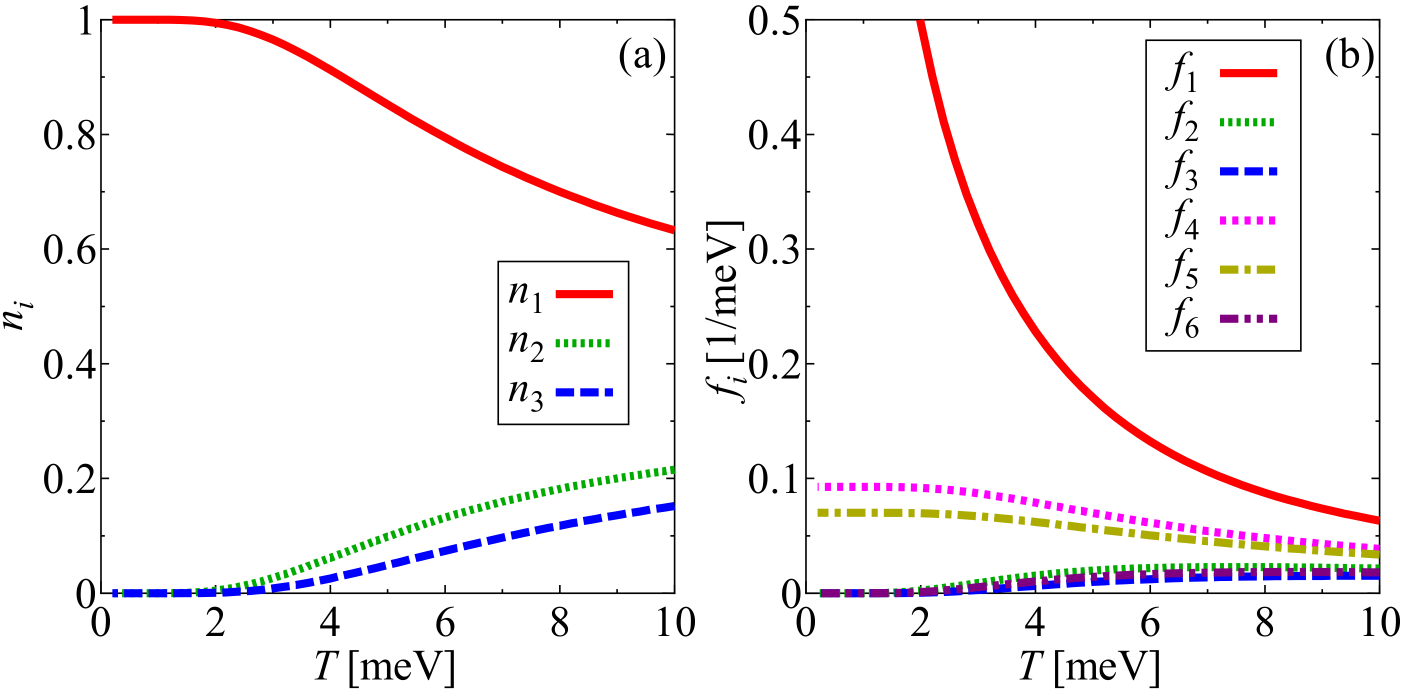}
\vspace{-0.3cm}
\caption{(Color online) 
$T$-dependence of (a) the particle-number $n_{i}=n_{i+}+n_{i-}$ for three KDs $(i=1,2,3)$ and (b) the functions $f_1$-$f_6$ in the single-site susceptibility $\chi_{m_1m_2m_3m_4}^{(0)}$.
}
\label{FigS1}
\vspace{+0.2cm}
\end{figure*}

The single-site susceptibility $\chi_{m_1m_2m_3m_4}^{(0)}$ for 6 CEF states in the experiment\cite{Nikitin2020} is given by the Lehmann representation as follows, 
\begin{align}
\chi_{m_1m_2m_3m_4}^{(0)}
&=\frac{1}{Z}\sum_{mm'}A_{mm'}^{m_1m_2}A_{m'm}^{m_4m_3}
\left\{\frac{\delta_{E_{m},E_{m'}}}{T}e^{-E_{m}/T}+\left(1-\delta_{E_{m},E_{m'}}\right)\frac{e^{-E_{m'}/T}-e^{-E_{m}/T}}{E_{m}-E_{m'}}\right\},
\end{align}
where $A_{mm'}^{m_1m_2}=\langle m|f_{m_1}^{\dagger}f_{m_2}^{}|m'\rangle$ and 
$f_{m}^{\dagger}$ is a creation operator for a $f$ electron of CEF state $m$ with energy $E_{m}$. 
$Z$ is a partition function $Z=2(1+e^{-\Delta_{1}/T}+e^{-\Delta_{2}/T})$ with the 1st~(2nd) excited level $\Delta_{1(2)}=10.78~(14.26)$~meV. The 6 CEF states of 3 KDs for $\{m,m',m_{i}\}$ are explicitly given by,
\begin{align}
&\Ket{1\pm}=\Ket{\Gamma_{7a\pm}}=\mp\sqrt{1-w^2}\ket{\pm\tfrac{5}{2}}\pm w\ket{\mp\tfrac{3}{2}},\label{eq:KD1}\\
&\Ket{2\pm}=\Ket{\Gamma_{7b\pm}}=w\ket{\pm\tfrac{5}{2}}+\sqrt{1-w^2}\ket{\mp\tfrac{3}{2}},\label{eq:KD2}\\
&\Ket{3\pm}=\Ket{\Gamma_{6\pm}}=\ket{\pm\tfrac{1}{2}},\label{eq:KD3}
\end{align}
where $w=0.95$\cite{Nikitin2020} and the particle-number $n_{i\pm}$ of each KD $|i\pm\rangle~(i=1,2,3)$ is also given by,
\begin{align}
&n_{1+}=n_{1-}=\frac{1}{Z}=\frac{1}{2(1+e^{-\Delta_{1}/T}+e^{-\Delta_{2}/T})},\\
&n_{2+}=n_{2-}=\frac{e^{-\Delta_{1}/T}}{Z}=\frac{e^{-\Delta_{1}/T}}{2(1+e^{-\Delta_{1}/T}+e^{-\Delta_{2}/T})},\\
&n_{3+}=n_{3-}=\frac{e^{-\Delta_{2}/T}}{Z}=\frac{e^{-\Delta_{2}/T}}{2(1+e^{-\Delta_{1}/T}+e^{-\Delta_{2}/T})}.
\end{align}

The single-site susceptibility $\chi_{m_1m_2m_3m_4}^{(0)}$ becomes finite only for certain sets of $m_1,m_2,m_3,m_4$ as summarized in Table~\ref{table-S2}, which results in the following form, 
\begin{align}
&\chi_{m_1m_2m_3m_4}^{(0)}=\delta_{m_3m_1}\delta_{m_2m_4}
\Bigl\{
\delta_{11}^{m_1m_2}f_1+\delta_{22}^{m_1m_2}f_2+\delta_{33}^{m_1m_2}f_3\nonumber\\
&\qquad\qquad\qquad\qquad
+\left(\delta_{12}^{m_1m_2}+\delta_{21}^{m_1m_2}\right)f_4
+\left(\delta_{13}^{m_1m_2}+\delta_{31}^{m_1m_2}\right)f_5
+\left(\delta_{23}^{m_1m_2}+\delta_{32}^{m_1m_2}\right)f_6
\Bigr\},\\
&\quad\delta_{ij}^{m_1m_2}=
\delta_{m_1,i+}\delta_{m_2,j+}
+\delta_{m_1,i-}\delta_{m_2,j-}
+\delta_{m_1,i+}\delta_{m_2,j-}
+\delta_{m_1,i-}\delta_{m_2,j+}
\quad(i,j=1,2,3),
\end{align}
where $f_{1}$-$f_{6}$ are functions with different $T$-dependence given in Table~\ref{table-S2}.

Figure~\ref{FigS1} shows (a) the number of occupied electrons $n_{i}=n_{i+}+n_{i-}$ in the three KDs and (b) the functions $f_1$-$f_6$ in the single-site susceptibility as a function of $T$ under the present CEF. 
The occupied number of the lowest KD $n_1$~(excited KDs $n_2,n_3$) increases~(decreases) with decreasing $T$, and finally $n_1\rightarrow 1~(n_2,n_3\rightarrow 0)$ for $T\lesssim 2~{\rm meV}$ as shown in Fig.~\ref{FigS1}(a). 
Corresponding to such the $T$-dependence of $n_{i}$, the Curie term of the lowest KD $f_{1}$ increases in proportion to $1/T$ with decreasing $T$ as shown in Fig.~\ref{FigS1}(b). 
The Van-Vleck terms between the lowest KD and the excited KDs, $f_4,f_5$, also increase with decreasing $T$, but they become constant at low temperatures of $T\lesssim 2~{\rm meV}$. 
The $T,O_{\Gamma}$-dependent differences in single-site multipole susceptibility shown in Fig.~2 of the main text are due to the presence or absence of $f_1$ and $f_4,f_5$ and the size of their coefficients, for example, the $f_1$ contribution of $H_0$ is larger than taht of $I$, resulting in the larger susceptibility of $H_{0}$ than that of $I$.

\begin{table}[t]
\caption{
Finite sets of the matrix elements of the single-site susceptibility $\chi_{m_1m_2m_3m_4}^{(0)}$.
}\label{table-S2}
\vspace{2mm}
\centering
\scalebox{1.00}{
\begin{tabular}{cc}
\hline
$\chi_{m_1m_2m_3m_4}^{(0)}$ &$m_1,m_2,m_3,m_4$ \\\hline\hline
$\displaystyle f_{1}\equiv\frac{1}{ZT}=\frac{n_{1\pm}}{T}$                     &\begin{tabular}{c} $1\pm,1\pm,1\pm,1\pm$ \\ $1\pm,1\mp,1\pm,1\mp$ \end{tabular}\\\hdashline
$\displaystyle f_{2}\equiv\frac{e^{-\Delta_{1}/T}}{ZT}=\frac{n_{2\pm}}{T}$ &\begin{tabular}{c} $2\pm,2\pm,2\pm,2\pm$ \\ $2\pm,2\mp,2\pm,2\mp$ \end{tabular}\\\hdashline
$\displaystyle f_{3}\equiv\frac{e^{-\Delta_{2}/T}}{ZT}=\frac{n_{3\pm}}{T}$ &\begin{tabular}{c} $3\pm,3\pm,3\pm,3\pm$ \\ $3\pm,3\mp,3\pm,3\mp$ \end{tabular}\\\hline
$\displaystyle f_{4}\equiv\frac{1-e^{-\Delta_{1}/T}}{Z\Delta_{1}}=\frac{n_{1\pm}-n_{2\pm}}{\Delta_{1}}$ &\begin{tabular}{c} $1\pm,2\pm,1\pm,2\pm$ \\ $1\pm,2\mp,1\pm,2\mp$ \\ $2\pm,1\pm,2\pm,1\pm$ \\ $2\pm,1\mp,2\pm,1\mp$ \end{tabular}\\\hdashline
$\displaystyle f_{5}\equiv\frac{1-e^{-\Delta_{2}/T}}{Z\Delta_{2}}=\frac{n_{1\pm}-n_{3\pm}}{\Delta_{2}}$ &\begin{tabular}{c} $1\pm,3\pm,1\pm,3\pm$ \\ $1\pm,3\mp,1\pm,3\mp$ \\ $3\pm,1\pm,3\pm,1\pm$ \\ $3\pm,1\mp,3\pm,1\mp$ \end{tabular}\\\hdashline
$\displaystyle f_{6}\equiv\frac{e^{-\Delta_{1}/T}-e^{-\Delta_{2}/T}}{Z(\Delta_{2}-\Delta_{1})}=\frac{n_{2\pm}-n_{3\pm}}{\Delta_{2}-\Delta_{1}}$ &\begin{tabular}{c} $2\pm,3\pm,2\pm,3\pm$ \\ $2\pm,3\mp,2\pm,3\mp$ \\ $3\pm,2\pm,3\pm,2\pm$ \\ $3\pm,2\mp,3\pm,2\mp$ \end{tabular}\\
\hline
\end{tabular}
}
\end{table}

\section{Derivation of RKKY interactions}
Here we derive the RKKY interaction $K_{m_1m_2m_3m_4}^{\alpha,\beta}(\bm{q})$ of Eq.~(3) in the main text. 
First, we introduce the generalized periodic Anderson~(GPAM) model\cite{Hanzawa2015} which can be written by,
\begin{align}
\mathscr{H}_{\rm GPAM}&=\mathscr{H}_{c\textrm{-}c}+\mathscr{H}_{c\textrm{-}f}+\mathscr{H}_{f\textrm{-}f}\label{eq:GPAM}\\
&=\sum_{\bm{k}\ell\ell'}h_{\ell\ell'}^{cc}(\bm{k})c_{\bm{k}\ell}^{\dagger}c_{\bm{k}\ell'}^{}
+\frac{1}{\sqrt{N}}\sum_{\bm{k}\ell}\sum_{\bm{i}\alpha m}\left(e^{-i\bm{k}\cdot\bm{R_i}}V_{\bm{k}\ell\alpha m}c_{\bm{k}\ell}^{\dagger}f_{\bm{i}\alpha m}^{}+h.c.\right)
+\sum_{\bm{i}\alpha}\mathscr{H}^{f}_{\bm{i}\alpha},
\end{align}
where the first term is the same in Eq.~(\ref{eq:HTB-k}) and the second term corresponds to the Fourier transform of the $f$-electron of the third term in Eq.~(\ref{eq:HTB-k}), where the sub-lattice degrees of freedom of Ce-$f$ electrons are explicitly written by $\alpha={\rm Ce1,Ce2}$. 
The third term is the local $f$-electron term including the onsite energy levels and Coulomb interactions for $f$ electrons at each Ce atoms, where we do not write them explicitly here. 

Next, we derive the effective Hamiltonian restricted to $f^{1}$ states by a second-order perturbation~(SOP) w. r. t. $\mathscr{H}_{c\textrm{-}f}$ in $\mathscr{H}_{\rm GPAM}$. 
The resulting Hamiltonian is the multi-orbital Kondo lattice model which is explicitly given by, 
\begin{align}
\mathscr{H}_{\rm MKL}&=
\sum_{\bm{i}\alpha m}\varepsilon_{m}^{f}f_{\bm{i}\alpha m}^{\dagger}f_{\bm{i}\alpha m}^{}
+\sum_{\bm{k}\ell\ell'}h_{\ell\ell'}^{cc}(\bm{k})c_{\bm{k}\ell}^{\dagger}c_{\bm{k}\ell'}^{}
+\sum_{\bm{i}\alpha mm'}\sum_{\bm{k}\bm{k'}\ell\ell'}J_{\bm{i}\alpha mm'}^{\bm{k}\ell,
\bm{k'}\ell'}f_{\bm{i}\alpha m}^{\dagger}f_{\bm{i}\alpha m'}^{}c_{\bm{k}\ell}^{\dagger}c_{\bm{k'}\ell'}^{},
\label{eq:MKL}
\end{align}
where $m$ represents 6 states of Eqs.~(\ref{eq:KD1})-(\ref{eq:KD3}) on each atom $\alpha$ in uc at $\bm{R_i}$ and $\varepsilon_{m}^{f}$ is the CEF energy level of the $f^1$ state given by $O(10^{0-1}~{\rm meV})$ as written in the main text, 
while the energy level of the $f^0$ state $\Delta_{0}$ is $O(10 ^{0}~{\rm eV})$ much larger than $\varepsilon_{m}^{f}$. Here we employ only the $f^0$-intermediate state for the Kondo coupling $J_{\bm{i}\alpha mm'}^{\bm{k}\ell,\bm{k'}\ell'}$ in Eq.~(\ref{eq:MKL}) and obtain as follows,
\begin{align}
J_{\bm{i}\alpha mm'}^{\bm{k}\ell,\bm{k'}\ell'}
=\frac{1}{N}V_{\bm{k}\ell\alpha m}^{}V_{\bm{k'}\ell'\alpha m'}^{*}
\left(
\frac{1}{\varepsilon_{\bm{k}\ell}^{c}+\Delta_{0}-\varepsilon_{m}^{f}}+
\frac{1}{\varepsilon_{\bm{k'}\ell'}^{c}+\Delta_{0}-\varepsilon_{m'}^{f}}
\right)
e^{-i(\bm{k}-\bm{k'})\cdot\bm{R_i}},
\end{align}
where the contribution of the $f^2$ intermediate states is treated as the same as that of the $f^0$-intermediate state. 
In a study of the dynamical mean field theory for CeB$_6$\cite{Otsuki2022}, the $f$-electron susceptibility has been calculated including the contribution of $f^n$-multiplet intermediate states on 1-ion, where the contribution of $f^2$ states is found to be less important for the main $\bm{q}$ dependence of the Stoner factor that determines the critical point of the multipole order. 
Therefore the present treatment taken only the $f^0$-contribution can give the proper $\bm{q}$-dependence of the RKKY interactions and obtain the dominant instability towards the multipole order. 
Moreover we fix the scattered $c$-electron energies to the Fermi energy $(=0)$, $\varepsilon_{\bm{k}\ell}^{c},~\varepsilon_{\bm{k'}\ell'}^{c}\rightarrow 0$. 
Then the Kondo coupling $J_{\bm{i}\alpha mm'}^{\bm{k}\ell,\bm{k'}\ell'}$ can be written by the following form, 
\begin{align}
&J_{\bm{i}\alpha mm'}^{\bm{k}\ell,\bm{k'}\ell'}
=\frac{1}{N}V_{\bm{k}\ell\alpha m}^{}V_{\bm{k'}\ell'\alpha m'}^{*}
\left(
\frac{1}{\Delta_{0}-\varepsilon_{m}^{f}}+
\frac{1}{\Delta_{0}-\varepsilon_{m'}^{f}}
\right)
e^{-i(\bm{k}-\bm{k'})\cdot\bm{R_i}}.
\end{align}

The RKKY Hamiltonian can be obtained from the SOP w. r. t. the Kondo coupling term together with the thermal average for the $c$ states. The final form is given by,
\begin{align}
&\mathscr{H}_{\rm RKKY}=-\sum_{\bm{i},\bm{\delta}}\sum_{m_1m_2}\sum_{m_3m_4}
K_{m_1m_2m_3m_4}^{\alpha,\beta}(\bm{\delta})
f_{\bm{i}\alpha m_1}^{\dagger}
f_{\bm{i}\alpha m_2}^{}
f_{\bm{i}+\bm{\delta}\beta m_4}^{\dagger}
f_{\bm{i}+\bm{\delta}\beta m_3}^{}
\label{eq:HRKKY-r},\\
&K_{m_1m_2m_3m_4}^{\alpha,\beta}(\bm{\delta})=\frac{1}{N}\sum_{\bm{q}}K_{m_1m_2m_3m_4}^{\alpha,\beta}(\bm{q})~e^{i\bm{q}\cdot\bm{\delta}},
\end{align}
where $K_{m_1m_2m_3m_4}^{\alpha,\beta}(\bm{\delta})$ is the RKKY coupling between $f$ states $(\alpha m_1m_2)$ at $\bm{R_i}$ and $(\beta m_3m_4)$ at $\bm{R_i}+\bm{\delta}$ with the primitive translation vector $\bm{\delta}$. The key quantity $K_{m_1m_2m_3m_4}^{\alpha,\beta}(\bm{q})$ is given by,
\begin{align}
&K_{m_1m_2m_3m_4}^{\alpha,\beta}(\bm{q})=C_{m_1m_2m_3m_4}\frac{1}{N}\sum_{nn'\bm{k}}
\mathcal{M}^{n\bm{k}}_{\beta m_3,\alpha m_1}
\mathcal{M}^{n\bm{k}+\bm{q}}_{\alpha m_2,\beta m_4}
\frac{f_{n'\bm{k}+\bm{q}}-f_{n\bm{k}}}{\varepsilon_{n\bm{k}}-\varepsilon_{n'\bm{k}+\bm{q}}},\\
&C_{m_1m_2m_3m_4}=\left(
\frac{1}{\Delta_{0}-\varepsilon_{m_1}^{f}}+\frac{1}{\Delta_{0}-\varepsilon_{m_2}^{f}}\right)\left(
\frac{1}{\Delta_{0}-\varepsilon_{m_3}^{f}}+\frac{1}{\Delta_{0}-\varepsilon_{m_4}^{f}}\right),\label{eq:C]}\\
&\mathcal{M}^{n\bm{k}}_{\alpha m,\beta m'}=\sum_{\ell\ell'}V_{\bm{k}\ell\alpha m}^{*}V_{\bm{k}\ell'\beta m'}^{}U_{n\bm{k}\ell}^{c*}U_{n\bm{k}\ell'}^{c},\label{eq:M}
\end{align}
where $C_{m_1m_2m_3m_4}$ is a energy denominator by the $f^{1}$-$f^{0}$ process and $\mathcal{M}^{n\bm{k}}_{\alpha m,\beta m'}$ denotes the scattering matrix element between the $f$-electron states $(\alpha m)$ and $(\beta m')$ via $c$-$f$ mixing with the $c$-electron band states with $(n\bm{k})$.
Here we set the excitation energy to $\Delta_{0}=2$~eV from the center of mass of the $f$ bands of LaCoSi shown in Fig.~1(a) of the main text, which yields the energy denominator $C_{m_1m_2m_3m_4}\simeq 4/\Delta_{0}^{2}=1~[{\rm eV^{-2}}]$ due to $\Delta_{0}\gg\varepsilon_{m}^{f}$, resulting in the same form of $K_{m_1m_2m_3m_4}^{\alpha,\beta}(\bm{q})$ as Eq.~(3) in the main text.

\section{Enhancement rates of the RPA susceptibilities}
Figure \ref{FigS2} shows the enhancement rates of the $\bm{q}=\bm{0}$ RPA susceptibilities for (a)-(d) uniform and (e)-(h) staggered components, excluding the TR-even $\Gamma_{1}$ mulitpoles. 
It can be seen that the uniform and staggered components exhibit similar behavior.
The enhancement rates for the quadrupoles and hexadecapoles have a small $T$-dependence and are hardly enhanced as shown in Figs.~\ref{FigS2}(a),(e) and Figs.~\ref{FigS2}(b),(f). These behavior can be explained by the $T$-dependence of the single-site susceptibilities in Fig.~2(a) and 2(b) of the main text, i.e., TR-even susceptibilities other than $\Gamma_{1}$ do not include the lowest KD Curie term contribution and are at most increased by the Van-Vleck term contribution. On the other hand, the rates for TR-odd multipoles of $(J_{z},~T_{z}^{\alpha},~D_{z}^{1\alpha,2\alpha})~[\Gamma_{2}]$ and $(J_{x/y},~T_{x/y}^{\alpha},~D_{x/y}^{1\alpha,2\alpha})~[\Gamma_{5}]$ enhance two to three times as shown in Figs.~\ref{FigS2}(c),(d) and (g),(h), which is attributed to the fact that the TR-odd susceptibilities shown in Figs.~2(c) and (d) of the main text include a Curie term contribution.

\begin{figure*}[t]
\centering
\includegraphics[width=12.0cm]{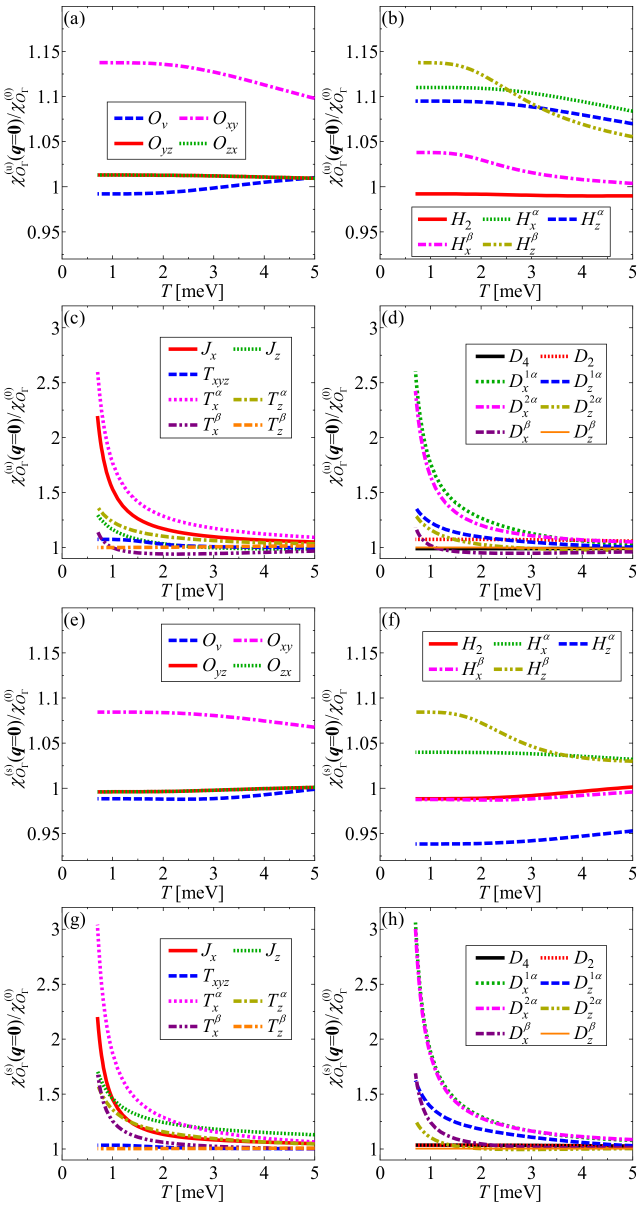}
\vspace{-0.3cm}
\caption{(Color online) 
The enhancement rates of the $\bm{q}=\bm{0}$ RPA susceptibility for (a)-(d) uniform and (e)-(h) staggered multipoles except for the TR-even $\Gamma_{1}$ multipoles as a function of $T$ for (a),(e) monopole and quadrupoles, (b),(f) hexadecapoles, (c),(g) dipoles and octupoles and (d),(h) triakontadipoles.
}
\label{FigS2}
\vspace{+0.2cm}
\end{figure*}
}

\end{document}